\documentclass[11pt]{article}

\usepackage[english]{babel}
\usepackage[utf8]{inputenc}

\usepackage{amssymb,dsfont,amsmath,amsfonts,hyperref,mdframed,mathrsfs,stmaryrd,bbold}
\usepackage[enableskew]{youngtab}
\usepackage[active]{srcltx}
\usepackage{slashed}
\usepackage{color}
\usepackage{graphicx}
\usepackage{url}

\advance \textheight by \topskip
\DeclareMathAlphabet{\mathpzc}{OT1}{pzc}{m}{it}
\usepackage{xcolor}

\usepackage[normalem]{ulem}
\usepackage{graphicx}
\usepackage{soul}
\usepackage{cite}

\usepackage{mathtools}

\usepackage{array}

\usepackage{cancel}

\definecolor{mygreen}{rgb}{0.1, 0.7, 0.2}

\newcommand{\rc}{\textcolor{red}}

\newcommand{\num}{N\textsuperscript{o}\,}

\def\be{\begin{equation}}
\def\ee{\end{equation}}
\def\bea{\begin{eqnarray}}
\def\eea{\end{eqnarray}}
\def\lpartial{\overleftarrow{\partial}}
\def\rpartial{\overrightarrow{\partial}}
\def\BB{\overleftrightarrow{B}}

\def\ie{\textit{i.e.} }
\def\eg{\textit{e.g.} }
\def\cf{\textit{cf.} }

\title{Pseudoclassical system with gauge and time-reparametrization invariance}

\author{Mauricio Valenzuela\\[12pt]
\textit{Facultad de Ingenier\'ia, Arquitectura y Dise\~no, Universidad San Sebasti\'an, Valdivia, Chile}\\
and \\
\textit{Centro de Estudios Cient\'{\i}ficos (CECs), Arturo Prat 514, Valdivia, Chile} 
}

\begin{document}

\maketitle

\begin{abstract}
We present a pseudoclassical mechanics model which exhibits gauge symmetry and time-reparametrization invariance. As such, first- and second-class constraints restrict the phase space, and the Hamiltonian weakly vanishes. We show that the Dirac conjecture does not hold---the secondary first-class constraint is not a symmetry generator---and only the gauge fixing condition associated with the primary first-class constraint is needed to remove the gauge ambiguities. The gauge fixed theory is equivalent to the Fermi harmonic oscillator extended by a boundary term. We quantize in the deformation quantization and in the Schr\"odinger representation approaches and observe that the boundary term prepares the system in the state of positive energy.   

 \end{abstract}

\tableofcontents

%%%%%%
\section{Introduction}
%%%%%%

Gauge theories are characterized by configuration spaces containing arbitrary functions of time. Because these systems are naturally non-deterministic additional constraints, \ie gauge fixing conditions, are needed to eliminate the dynamical ambiguities. The prescription to do so is provided by the constrained Hamiltonian analysis introduced by Dirac  \cite{Dirac2001,Henneaux:1992ig}.  The standard procedure, as it is usually taught, includes the strong assumption of Dirac's conjecture \cite{Dirac2001}, which says that both primary and   secondary first-class constraints are gauge symmetry generators.  

Dirac---in his own words---did not know counterexamples for his conjecture \cite{Dirac2001}. However, in 1979 Cawley provided one \cite{Cawley1979} and since then many counterexamples have been found \cite{Henneaux:1992ig,Pons:2004pp,Earman:2003cy}. 

The Dirac conjecture implies---according to the algorithm every gauge symmetry generator should be accompanied by a gauge fixing condition---the elimination of the degrees of freedom conjugate to the constrained phase space directions.

Consequently, wrongly assuming the Dirac conjecture may lead to an incorrect counting of degrees of freedom, and possibly to an unintended truncation of the model at hand. 

It has been recently proposed that this may be the case in the massless Rarita-Schwinger action \cite{Valenzuela:2022gbk}, which raises a fundamental question about the real field content of supergravity: Is the spin-$\frac12$ a pure gauge mode of simple supergravity \cite{fnf,dz}? It has been known for a while that the spin-half projection of the massless Rarita-Schwinger is described by the Dirac action---this is the essence of the \textit{unconventional supersymmetry} approach \cite{Alvarez:2011gd,Alvarez:2020izs,Alvarez:2021zhh,Alvarez:2020qmy,Alvarez:2021qbu}---which supports an answer to that question in the negative. What was shown in \cite{Valenzuela:2022gbk} is that the conclusion that the spin-half sector of supergravity is ``pure gauge" follows from assuming the Dirac conjecture, which would demand a gauge fixing conditions in addition to the secondary first-class constraint. This implies the elimination of one of the two spin-half Poincar\'e irreps present in the vector spinor. If the Dirac conjecture is not assumed, the massless Rarita-Schwinger equations in the standard vanishing gamma-trace gauge (conjugate to the primary first-class constraint) decouple into spin-$\tfrac{3}{2}$ and  spin-$\tfrac{1}{2}$ components, and an explicit solution containing both spins can be written \cite{Valenzuela:2022gbk}.

Since the Dirac conjecture cannot be taken for granted and since it affects the counting of degrees of freedom, its correct understanding becomes extremely important. Thus, the construction of Dirac conjecture counterexamples is relevant, either for pedagogical reasons or for research purposes. This is the goal of this article. 

More precisely, the model to be discussed here is of the \textit{pseudoclassical mechanics} type \cite{Berezin:1976eg,Casalbuoni:1976tz,Casalbuoni:1975bj}---classical mechanics with anti-commuting variables---which exhibits fermionic gauge symmetry and time-reparametrization invariance. In the Hamiltonian formulation, the model possesses first-class and second-class  primary constraints, and one first-class secondary constraint.

It has been argued \cite{Henneaux:1992ig} that the counterexamples of the Dirac conjecture may have quantization problems, at least in some cases. Thus, in order to show that this is not the case here,  we shall consider two quantization approaches. In the first case, we map classical functions to operators and find the Hilbert space. In the second, we employ the deformation quantization scheme started by Weyl, Wigner, Groenewold, and Moyal \cite{Weyl:1927vd,Wigner:1932eb,Groenewold:1946kp,Moyal:1949sk,Bayen:1977ha,Bayen:1977hb,Curtright:2014sli}, adapted to Fermi variables \cite{Fairlie:1989vv,Zachos:1999mp} (see \cite{Hirshfeld:2002ki,Galaviz:2006ni,Allen:2015lma} for applications).

We shall see that our model is equivalent to the Fermi oscillator \cite{DeWitt:2012mdz} with the secondary first-class constraint set by a boundary term. In the quantum setting the secondary first-class constraint sets the Fermi oscillator in a stationary state of positive energy.

\section{Counting degrees of freedom  with and without the Dirac conjecture}

Recall that the constraints in the Hamiltonian form of a system are classified as: \textit{primary constraints},  being the ones necessary to invert the Legendre transform; \textit{secondary constraints}, appearing as consistency conditions necessary to preserve the primary constraints; and as \textit{first-class constraints} (FCC) and \textit{second-class constraints} (SCC), depending on their algebra with respect to the Poisson bracket product \cite{Dirac2001}. 

In his \textit{Lecture on quantum mechanics}, Dirac stated that
``I think it may be that all the
first-class secondary constraints should be included among the transformations which don't change the
physical state, but I haven't been able to prove it. Also, I
haven't found any example for which there exists first-class
secondary constraints which do generate a change in the
physical state"

Though it is clear that first-class primary constraints are gauge symmetry generators, the Dirac conjecture implies that secondary constraints should also be regarded as such. According to this logic, all first-class constraints must be accompanied by gauge fixing conditions, hence:
 \begin{center}
\num gauge fixing  conditions
=
\num  1st class constraints.
\end{center}
Thus the number of phase space directions that are removed is twice the number of FCC.  It follows that \textit{the gauge strikes twice}. However, if the Dirac conjecture does not hold, only primary FCC need gauge fixing conditions, and the secondary FCCs do not strike a second time.

The test of the Dirac conjecture consists of verifying whether or not the equations of motion are invariant under the transformation generated solely by secondary FCC. Another way to test the Dirac conjecture is to verify if imposing only gauge fixing conditions for primary FCC the system becomes deterministic, \ie free of gauge ambiguities. Recall that gauge fixing conditions are external inputs imposed, by hand, to remove the arbitrary functions of time which make the system's dynamics non-deterministic. Hence, it is reasonable to minimize the number of external inputs, imposing only those that are necessary to obtain a deterministic result.

Castellani's  formal analysis \cite{Castellani:1981us} shows that secondary FCCs do not generate independent gauge symmetries, but they are needed as part of the true gauge generator,
\be\label{Castellani}
G=\sum_{n=0}^k \epsilon^{(k-n)}G_n\,,\qquad \epsilon^{(l)}:=\frac{d^{l}\epsilon}{dt^{l}}\,,
\ee
here referred to as \textit{Castellani chain}. This a linear combination starting with primary FCCs, $G_0$, followed by secondary FCC, $G_1$, and subsequent descendants for $k>1$, with coefficients given by the time-derivatives of the gauge parameters $\epsilon$. 

The number of independent Castellani chains can be less than the total number of FCCs since secondary FCC are discounted as independent symmetry generators. Hence,  the number of gauge fixing conditions---necessary to intersect the independent gauge orbits---may be less than the total number of FCC. In those cases, the assumption of the Dirac conjecture might lead to imposing unnecessary external constraints misinterpreted as gauge fixing conditions, which remove physical degrees of freedom instead of pure gauge modes. We shall see next that this is the case in the model introduced next.

%%%%%%%%%%%%%%%%%%%%%%%%%%%%%%%%%%%%%%%%%
\section{Lagrangian formulation}\label{sec:model}
%%%%%%%%%%%%%%%%%%%%%%%%%%%%%%%%%%%%%%%%%

Consider the following system, 
\be\label{L}
L= i \left(\theta \zeta - \dot\lambda \zeta \right),
\ee
for three real Grassmann-odd anticommuting functions of time $\theta(t),\,\lambda(t),\,\zeta(t)$.
 
The action can be spelled out in the form,
\be\label{S}
S= i \int \left(\Omega \zeta - d\lambda \, \zeta\right),
\ee
where $\Omega:=dt\theta$ and $d=dt\, \partial_t$, is explicitly invariant regarding time reparametrizations, 
\be\label{rep}
t\:\longrightarrow\: t'(t)
,\qquad  \theta \:\longrightarrow\:  \theta'=\Big(\frac{\partial t'}{ \partial t}\Big)^{-1} \theta\,,
\ee
while $\zeta$ and $\lambda$ do not transform.
 
The Euler-Lagrange equations are given by:
\be\label{ELeq}
\dot \lambda-\theta=0\,, \qquad \zeta=0\,,\qquad \dot\zeta=0 \,.
\ee
Hence $\zeta$ is auxiliary. Equations \eqref{ELeq} are left invariant with respect to the time reparametrization \eqref{rep} up to $\partial t'(t)/\partial t\neq 0$ factors.

The system has the gauge symmetry,
\be\label{gauge}
\delta\theta=\dot\epsilon\,,\qquad \delta\lambda=\epsilon\,,\qquad \delta\zeta=0\,.
\ee

We can fix the gauge imposing the relation:
\be\label{gf}
\theta+i\omega\lambda\approx 0\,.
\ee
It describes a one-dimensional subspace in the $(\lambda,\theta)$ plane.  The gauge is reachable since one can perform a transformation with the parameter 
\be
\epsilon=-\int ds \, \Theta(t-s) e^{-i\omega (t-s)} (\theta(s)+i\omega\lambda(s))\,,
\ee
where $\Theta$ is the Heaviside function, such that  \eqref{gf} is satisfied.

Going back to \eqref{ELeq}, the gauge fixed $\theta$ yields
\be\label{waveq}
\dot \lambda+i\omega\lambda=0\,,
\ee
and the solution is given by,
\be\label{sol}
\lambda(t,\omega)=\lambda_0\,  e^{-i\omega t}\,,
\ee
with $\lambda_0$ an anticommuting constant. 

A different gauge choice, with $\theta+i\omega'\lambda\approx 0$ yields equivalent physics, since
\be
\dot \lambda+i\omega'\lambda=\frac{w'}{w} \left( \frac{w}{w'}\frac{d \lambda}{dt}+i\omega\lambda\right)
\ee
amounts to reparametrization of time in \eqref{gf} with lapse function 
\be
\frac{\partial t}{\partial t'} =\frac{\omega}{\omega'}.
\ee
Therefore we can always set $\omega$ constant. Once the function $\omega$ is specified, both, gauge symmetry and the time reparametrization invariance cease. 
The freedom in the choice of $\omega$ does not represent a redundancy of the degrees of freedom; it just reflects the freedom on the definition of time.

\section{Hamiltonian formulation}\label{sec:Hamilton}

The Hamiltonian description of the system \eqref{L} occurs in the phase space of canonical  variables $(\theta,\pi_\theta;\,\lambda,\pi_\lambda;\,\zeta,\pi_\zeta)$.

The Legendre transform,
\be
\pi_\theta=\frac{\partial L}{\partial \dot \theta}\approx 0\,,\qquad \pi_\lambda=\frac{\partial L}{\partial \dot \lambda}\approx 0\,,\qquad \pi_\zeta=\frac{\partial L}{\partial \dot \zeta}=  - i \lambda\,,
\ee
produces  the \textit{primary constraints} 
\be\label{pconst}
\pi_\theta\approx 0\,,\qquad \chi_1=\pi_\zeta \approx 0\,,\qquad \chi_2=\pi_\lambda+ i\zeta \approx 0\,,
\ee
which describes the phase space sub-manifold containing the physical degrees of freedom. 

The unconstrained phase space comes with the Poisson bracket,
\bea
\{f,g\}&=& (-1)^{|f|} \left(\frac{\partial f}{\partial \theta} \frac{\partial g}{\partial \pi_\theta}+ \frac{\partial f}{\partial \pi_\theta}\frac{\partial g}{\partial \theta}  +\frac{\partial f}{\partial \lambda} \frac{\partial g}{\partial \pi_\lambda}+\frac{\partial f}{\partial \pi_\lambda}\frac{\partial g}{\partial \lambda} +\frac{\partial f}{\partial \zeta} \frac{\partial g}{\partial \pi_\zeta}+\frac{\partial f}{\partial \pi_\zeta} \frac{\partial g}{\partial \zeta}\right)\,,
\eea
where $|f|=0,1,$ is the even, odd, Grassmann parity of the function $f$.

It turns out that the constraints $\chi_i\,,i=1,2$ are of second-class, 
\be
\{\chi_i,\chi_j\}= C_{ij}\,,\qquad C_{ij}:= -i \left(
\begin{tabular}{cc}
0&1\\
1&0
\end{tabular}\right),
\ee
since  $C_{ij}$ is invertible,
\be\label{Ci}
C^{-1\, ij}=i \left(
\begin{tabular}{cc}
0&1\\
1&0
\end{tabular}\right)\,.
\ee
The constraint $\pi_\theta\approx 0$ is first-class since $\{\pi_\theta,\chi_j\} \approx 0$.

The canonical Hamiltonian,  $H_0=\dot\theta \pi_\theta +\dot\lambda \pi_\lambda +\dot\zeta \pi_\zeta - L $,
$$
H_0= -i \theta \zeta,
$$
must be complemented with the primary constraints,
\be\label{HT}
H_T:= H_0 +\pi_\theta \nu + \chi_i \mu^i  \,,
\ee
which defines the \textit{total Hamiltonian}. Here  $\nu$ and $\mu^i$ are Lagrange multipliers.
In Hamiltonian form, the action principle is equivalent to 
\bea\label{Haction}
S_H&=&\int dt ( \dot\theta \pi_\theta +\dot\lambda \pi_\lambda +\dot\zeta \pi_\zeta - H_T)\,,
\eea
which is stationary for field configurations satisfying differential equations equivalent to \eqref{ELeq}, and we recover the Lagrangian action \eqref{S} on the surface of the primary constraints.  Indeed, the variational principle \eqref{HT} yields 
\bea \rc{}
&\dot \lambda=-\mu^2\,,\qquad \dot\zeta=-\mu^1\,,\qquad \pi=0\,,&\label{eqlzt}
\\& \dot\theta=-\nu\,.&\label{eqnu}
\eea
The evolution of the classical observables is given by,
\be 
\nonumber
\dot f =\partial_t f + \{f,H_T\}\,,
\ee
which produces the consistency conditions,
\bea
&\dot \pi_\theta=i\zeta \approx 0\,,&\label{sec}\\
&\dot \chi_1=0\,\,\Rightarrow \,\, \mu^2=-\theta \,,\qquad \dot \chi_2=0\,\,\Rightarrow \mu_1=0\,.&\label{eqchi}
\eea 

From  \eqref{eqlzt}, \eqref{sec} and \eqref{eqchi},  we recover the Euler-Lagrange equations \eqref{ELeq}, while 
$\nu$ and $\theta$ should still be determined.

\subsection*{Gauge symmetry generator}

The gauge symmetry transformation of the Hamiltonian system of equations \eqref{eqlzt}-\eqref{eqchi}, and of the action \eqref{Haction}, is generated by \eqref{gauge} together with, 
\be\label{gauge2}
\delta\nu=-\ddot\epsilon\,,\qquad \delta\mu^2=-\dot\epsilon\,\,.
\ee

Note that the secondary constraint $\zeta\approx 0$ and $\chi_2$ imply that $\pi_\lambda\approx 0$. Thus we obtain two first-class constraints 
\be\label{1stcc}
\pi_\theta\approx0,\qquad \pi_\lambda\approx 0\,,
\ee
since they have a trivial Poisson bracket with $\chi_i$ and among themselves.

None of the constraints \eqref{1stcc} generate independent symmetries of the action \eqref{Haction}, and of the field equations. Indeed, it can be verified that with different parameters $\epsilon$, $\epsilon'$,
\be\label{G1}
G(\epsilon,\epsilon')= \epsilon'\, \pi_\theta +\epsilon\, \pi_\lambda\,,
\ee
the variation 
\be
\delta f = \{f,G(\epsilon,\epsilon')\},
\ee
does not generate any symmetry. We must set $\epsilon'=\dot \epsilon$ complemented with \eqref{gauge2} to obtain a true symmetry generator, which turns out to be the Castellani chain \cite{Castellani:1981us}, 
\be\label{G2}
G(\epsilon)= \dot\epsilon\, \pi +\epsilon\, \pi_\lambda\,.
\ee
Hence $\pi_\lambda$ is not a symmetry generator, and the Dirac conjecture does not hold. 
%We shall confirm this by the gauge fixing procedure.

\section{Reduction and gauge fixing}\label{sec:red}

First, let us pass to the phase space submanifold defined by the SCC $\chi_i=0$ \eqref{pconst}. There the momentum variables $\pi_\zeta$ and $\zeta$ can be removed and the Dirac brackets $\{f,g\}_D:= \{f,g\} - \{f,\chi_i\} \,C^{-1\,ij}\{\chi_j,g\}$ reduces to 
\be
\{f,g\}_D= ( - 1)^f \left(\frac{\partial f}{\partial \theta} \frac{\partial g}{\partial \pi_\theta}+ \frac{\partial f}{\partial \pi_\theta}\frac{\partial g}{\partial \theta} +\frac{\partial f}{\partial \lambda} \frac{\partial g}{\partial \pi_\lambda}+\frac{\partial f}{\partial \pi_\lambda}\frac{\partial g}{\partial \lambda} 
\right)\,,
\ee
on functions of $f(\theta,\pi_\theta;\lambda,\pi_\lambda)$. Consequently, the relevant canonical relations are given by, 
 \be\label{alg}
\{\theta,\pi_\theta\}_D=-1\,,\qquad  \{\lambda,\pi_\lambda\}_D=-1.
\ee
Then the reduced Hamiltonian reads,
\be\label{rH}
H_R:= \theta \pi_\lambda +\pi_\theta \nu\,.
\ee
This Hamiltonian yields equations of motion 
\bea\label{eomf}
\frac{d f}{dt} &=& \frac{\partial f}{\partial t} - \theta \frac{\partial f}{\partial \lambda} +\nu \frac{\partial f}{\partial \theta}\,,\qquad \pi_\theta \approx 0\,,\quad \pi_\lambda \approx 0\,,
\eea
which yields \eqref{eqlzt}-\eqref{eqnu} with Lagrange multipliers \eqref{eqchi} $\mu ^1=0$, $\mu^2=-\theta$. 

The apparently undetermined functions of time are given by $\theta$, $\nu$, and $\lambda$. However, there are two linear equations relating them,
\be\label{ltn}
\dot \lambda-\theta=0\,,\qquad \dot\theta+\nu = 0\,,
\ee
and hence adding one more constraint suffices in order to determine the system completely. The missing equation is the gauge condition associated with the primary first-class constraint $\pi_\theta\approx 0$, a restriction on $\theta$.  We choose the gauge condition \eqref{gf},
with constant $\omega$. It follows from \eqref{ltn} and \eqref{gf} that $\lambda$ must satisfy the wave equation \eqref{waveq}, whose solution is given by \eqref{sol}, and $\nu=i\omega \theta= \omega^2 \lambda$ ceases to be arbitrary. Therefore, there is no need to add a new external condition conjugate to the secondary FCC $\pi_\lambda\approx 0$ in order to obtain a deterministic set of equations.

Note that the most general gauge fixing consists of $\theta$ expanded as a function of $(\lambda,\pi_\lambda,\pi_\theta)$. However, the terms containing $\pi_\lambda\approx0\approx \pi_\theta$ weakly vanish and can be discarded. Thus the expansion of $\theta$ reduces to a linear function of $\lambda$.

Setting the constraints $\pi_\theta=0$ and $\theta=i\omega \lambda$ strongly, the Dirac bracket is reduced to,
\be\label{rDB}
\{f,g\}_R= (-1)^f \left( \frac{\partial f}{\partial \lambda} \frac{\partial g}{\partial \pi_\lambda}+\frac{\partial f}{\partial \pi_\lambda}\frac{\partial g}{\partial \lambda} 
\right)\,.
\ee
The gauge-fixed Hamiltonian corresponds to the Fermi oscillator,  
\be\label{Hclas}
H_{fix}=-i\omega\lambda \pi_\lambda\,,
\ee
and the observables evolve according to, 
\be\label{eomeff2}
\dot f=  - i\omega  \lambda \frac{\partial f}{\partial \lambda}\,,\qquad  \pi_\lambda\approx 0\,.
\ee

We conclude that the Dirac conjecture does not apply to our model, since the secondary first-class constraint $\pi_\lambda$ does not generate an independent gauge symmetry, and since a single gauge fixing condition  suffices in order to determine all arbitrary functions of time present in the system. Note that the wrong assumption of the Dirac conjecture would imply the elimination of $\lambda(t)$, since this is the conjugate of the secondary first-class constraint $\pi_\lambda\approx 0$.

\subsection{The secondary first-class constraint as an initial condition}

Since the equations of motion of $\lambda$ and $\pi_\lambda$ are first-order, only one boundary condition is enough to determine the integration constants. This suggests that the gauge fixed  theory can be expressed with the first-class constraint $\pi_\lambda\approx 0$  imposed as a boundary condition, instead of writing it together with the equations of motion as in \eqref{eomeff2}. 

This is, without the constraint  $\pi_\lambda\approx 0$, we would find that the equations of motion, obtained from $\dot f =\{f,H_T\}$,
\be\label{lpeom}
\dot\lambda+i\omega \lambda=0\,,\qquad \dot\pi_\lambda-i\omega \pi_\lambda=0\,.
\ee
have general solutions,
\be
\lambda(t)=\lambda_\circ \,e^{-i\omega t}\,,\qquad \pi_\lambda(t)=\pi_{\lambda\, \circ} \, e^{i\omega t},
\ee
with  integration constants $(\lambda_\circ,\pi_{\lambda\, \circ})$. Setting 
\be
\lambda(0)=\lambda_\circ\,,\qquad \pi_{\lambda\, \circ}(0)=0\,,
\ee
for $\lambda_\circ\neq0$, we would obtain the same result as in \eqref{eomeff2}, since for all $t$ we have that $\pi_\lambda(t)=0$.

This suggests the following effective theory,
\bea\label{Seff}
S_{eff}&=&\int_{t_i}^{t_f}  dt \, ( \dot\lambda \pi_\lambda + i \omega \lambda \pi_\lambda) + \rho_f \pi_{\lambda}(t_f)\,,
\eea
is equivalent to \eqref{Haction}. 

In the interval $(t_i,t_f)$, the variation of the action vanishes if the field equations \eqref{lpeom} are satisfied. At the ends of the intervals the boundary term,
\be
\delta \lambda(t_f) \pi_\lambda(t_f)-\delta \lambda(t_i) \pi_\lambda(t_i) + \delta \rho_f  \pi_\lambda(t_f)  + \rho_f  \delta  \pi_\lambda(t_f) =0 \,,
\ee
must also vanish. It follows that for given non-vanishing initial condition $\lambda(t_i)$,  $\delta \lambda(t_i)=0$,  $\pi_\lambda$  must vanish at $t=t_f$ (from the variation of $\rho_f$) and consequently $ \pi_{\lambda}(t)=0$ for all times. The evolution of $\lambda(t)$ is determined by its field equation and initial value. The role of the boundary term $\rho_f \pi_{\lambda}(t_f)$  is to enforce the boundary condition $ \pi_{\lambda}(t_f)=0$, by variation of the Lagrange multiplier $\rho_f$. Since the system first-order, we cannot simultaneously fix $\lambda(t)$ at both extremes, $t=t_i$ and $t=t_f$.  Thus $\lambda(t_f)$ is free and $\pi_\lambda(t_f)=0$.

\section{Quantization}\label{sec:Qn}

There are many counterexamples to the Dirac conjecture. It is argued \cite{Henneaux:1992ig}, however, that without the assumption of the Dirac conjecture the quantization of those systems may be inconsistent.  We shall see that in the system proposed here, this is not the case.

We shall quantize in two different frameworks: quantization in Hilbert space and deformation quantization. In the first case we use two alternative representations, in terms of matrix operators, and as differential operators on Grassmann variables (Schr\"odinger realization).  The treatment of the Fermi oscillator to be considered here is close to references \cite{Hirshfeld:2002ki,Galaviz:2006ni} in deformation quantization, and to \cite{Allen:2015lma} in the operator approach.  It turns out that the model is equivalent to the Fermi oscillator prepared in the positive energy state.

\subsection{Operator correspondence}

We start by postulating a correspondence between the classical variables and operators: $(\lambda,\pi_\lambda)\rightarrow (\hat \lambda,\hat \pi_\lambda)$, while the Poisson bracket is mapped to a $\mathbb{Z}_2$ graded commutator, 
\be\label{qnpfg}
 \widehat{ \{f,g\}}_{D} = \frac{{[\hat f,\hat g]}}{i\hbar} \,,\qquad [\hat f,\hat g ]:= \hat f \hat g -(-1)^{|f||g|} \hat g\hat f\,.
\ee
Thus we should look for representations of the algebra, 
\be
[\hat \lambda,\hat \pi_\lambda ]= -i\hbar\,.
\ee
Modulo ordering,  the classical Hamiltonian \eqref{Hclas} is mapped to the quantum operator,
\be\label{hatH}
\hat H=-\frac{i\omega}{2} (\hat \lambda \hat \pi_\lambda - \hat \pi_\lambda \hat \lambda) \,.
\ee

The Schr\"odinger equation is then,
\be\label{tdsc}
i\hbar \frac{d}{dt} |\phi(t)\rangle - \hat H |\phi(t)\rangle=0\,,
\ee
and the stationary Schr\"odinger equation, 
$$
\hat H |\psi_E\rangle=E |\psi_E\rangle\,,
$$
is obtained by separation of the time coordinate: $|\phi(t)\rangle:=e^{ -i \frac E \hbar t} |\psi_E\rangle$.

The Hamiltonian \eqref{hatH} corresponds to the Fermi oscillator \cite{Hirshfeld:2002ki,Galaviz:2006ni,DeWitt:2012mdz}, which has two energy levels.
Hence the wave function can be expanded as,
\be\label{tdep}
|\phi(t)\rangle = e^{- i \frac \omega 2  t} |\psi_+\rangle +  e^{ + i \frac\omega 2 t} |\psi_-\rangle\,,
\ee
and $|\psi_+\rangle$ and $|\psi_-\rangle$ are initial states. As we shall see, 
the latter are energy eigenstates with the spectrum,
\be\label{spec}
\hat H |\psi_\pm\rangle=E_\pm |\psi_\pm\rangle\,,\qquad E_\pm  = \pm \frac{\omega \hbar}2\,.
\ee 

We shall also verify that the quantum analogue of the constraint $\pi_\lambda\approx 0$  can be implemented in two equivalent ways:
\be\label{qmc}
\hat \pi_\lambda|\phi(t) \rangle=0\,,\qquad\hbox{or}\qquad \hat \pi_\lambda|\phi(t_f) \rangle=0\,,
\ee
namely for all $t$,  or as the quantum analogue of the final-time condition \eqref{Seff}. In the latter case, we say that the fermion is prepared in a vanishing momentum state at $t=t_f$.

\subsubsection*{Matrix representation}

In a matrix representation, the operator correspondence can be given in the form,
\be
\hat \lambda=
\left(
\begin{array}{cc}
 0 &  0  \\
1 & 0  
\end{array}
\right)
 ,\qquad  \hat \pi_\lambda= 
\left(
\begin{array}{cc}
 0 &   -i\hbar \\
  0 & 0  
\end{array}
\right)
,\qquad  \hat H= \frac{\omega \hbar}2 \left(
\begin{array}{cc}
 1 &  0 \\
  0 & -1  
\end{array}
\right)
\ee
This representation is compatible with the reality conditions,
\be\label{hc}
\hat{\bar{\lambda}}=\frac{i}\hbar\hat\pi_\lambda\,,\qquad \hat{\bar{\pi}}=i\hbar\hat\lambda\,,\qquad  \hat H^\dagger=\hat H\,,
\ee
where $\hat{\bar{\lambda}}:=\hat\lambda^\dagger$ and  $\hat{\bar{\pi}}:=\hat\pi_\lambda^\dagger$.

The  Hamiltonian eigenstates  \eqref{spec} are given by
\be\label{est}
 |\psi_+\rangle = \left(\begin{array}{c}
 \psi_+ \\
0 
\end{array}
\right)\,,\qquad |\psi_-\rangle = \left(\begin{array}{c}
0 \\
 \psi_- 
\end{array}
\right)\,.
\ee 
 The condition  \eqref{qmc} can be implemented as $\hat \pi_\lambda |\phi(t_f)\rangle$, which implies $\psi_-=0$, and the system stays in the $E_+$ eigenstate for all values of the time $t$. 

\subsubsection*{Schr\"odinger realization }

Now the operator correspondence (\cf \cite{Allen:2015lma}) is given by,
\be
\hat \lambda= \lambda
 ,\qquad  \hat \pi_\lambda= -i\hbar\frac{\partial}{\partial\lambda}
,\qquad  \hat H= \frac{\hbar\omega }2 \left(\frac{\partial}{\partial\lambda} \lambda - \lambda \frac{\partial}{\partial\lambda}
\right).
\ee
The Eigenstates of the Hamiltonian are spanned in terms of the $\lambda$ variable as $|\psi\rangle=\psi_+ + \lambda \psi_-$, where $\psi_\pm$ are real commuting variables. The inner product is given by an integral formula,
\be\label{scalar}
\langle \bar\psi|\phi\rangle=\int \frac{d\lambda d\bar\lambda}{\mathcal{N}} (\bar\psi_+ +  \bar\psi_- \bar\lambda)(\phi_+ + \lambda \phi_-) = \frac{1}{\mathcal{N}} (\bar\psi_+ \phi_++  \bar\psi_-\phi_-)\,.
\ee
where $\mathcal{N}=|\psi_+|^2+|\psi_-|^2$ is a normalization constant, and we have used the standard rules for Berezin integrals.
Now we obtain the spectrum \eqref{spec} with 
\be
|\psi_+\rangle = \psi_+\,,\qquad |\psi_-\rangle = \lambda \psi_-\,,
\ee 
and the time-dependent solutions of the Sch\"rodinger equations are given as in \eqref{tdep}. Once again, the constraints \eqref{qmc} imply $\psi_-=0$. The operators $(\hat\lambda,\hat\pi_\lambda)$ satisfy the conjugation relations \eqref{hc} with respect to scalar product \eqref{scalar}.

\subsection{Deformation quantization approach}

In the deformation quantization approach, the classical functions remain invariant but the product of the canonical variables is deformed, 
\be
f\,g\quad \rightarrow \quad f*g=fg + O(\hbar)\,,
\ee
where the star-product  $*$, to be specified,  is equivalent to an expansion in $\hbar$ around $fg$.

The Poisson bracket is then deformed,
\be
 \widehat{ \{f,g\}}_{D} = \frac{{[\hat f,\hat g]_*}}{i\hbar}\,,\qquad  [f,g]_* = f*g -(-1)^{fg} g* f\,,
\ee
and it is such that in  the classical limit 
\be\label{class}
\lim_{\hbar\rightarrow 0} \frac{[f,g]_*}{i\hbar}=\{f,g\}\,.
\ee
it becomes, literally, the Poisson bracket.

We define the star-product
\be\label{star}
f* g := f \exp\Big(\frac{\hbar}{2i} \BB  \Big) g\,,
\ee
by the exponentiation of the Poisson bi-vector,
\be
\BB:= \frac{\lpartial}{\partial \lambda} \frac{\rpartial}{\partial \pi_\lambda}+ \frac{\lpartial}{\partial \pi_\lambda} \frac{\rpartial}{\partial \lambda} \,,
\ee
in the convention $f\lpartial=-(-1)^{|f|} \rpartial f$, for $\partial=\partial_\lambda,\partial_{\pi_\lambda}$. 

Since $\BB^3=0$, the expansion of the exponential \eqref{star} contains up to order 
\be
\BB^2=2  \frac{\lpartial}{\partial \lambda} \frac{\lpartial}{\partial \pi_\lambda}   \frac{\rpartial}{\partial \pi_\lambda} \frac{\rpartial}{\partial \lambda} \,.
\ee

In this framework, the Moyal equation 
\be\label{Moyal}
i\hbar \frac{dW}{dt}=i\hbar \frac{\partial W}{\partial t} + [W,H]_*=0\,,
\ee
provides the relevant distribution, namely the Wigner function. The classical limit of the Moyal equation is the Liouville theorem for the phase space distribution. 

The most general form of $W$ is,
\be\label{W1}
W(t)=w_0 + \lambda\, w_1\ +  \pi_\lambda\, w_2 + \lambda\pi_\lambda\, w_3\,,
\ee
where $w$'s are functions of time. 
The mean value of an observable is given by,
\be\label{mean}
\langle f \rangle = \int \frac{ d\pi_\lambda d\lambda}{\mathcal{N}}  \; f*W\,.
\ee
We demand $W$ to be Grassmann-even in order to preserve the Grassmann parity of the mean values \eqref{mean}, hence  ${|f|}=(0,1,1,0)$ for $f=(w_0,w_1,w_2,w_3)$ respectively. Since $H$ commutes with $w_0$ and $\lambda\pi_\lambda\, w_3$ the coefficients $w_1$ and $w_3$ must be time-independent $\partial_tw_0=0=\partial_t w_3$, in order to satisfy \eqref{Moyal} and the remaining coefficients must have the form,
\be
w_1 =\tilde w_1 \, \exp\left(-i\frac{w}\hbar t\right)\,,\qquad  w_2=\tilde w_2 \,\exp\left(i\frac{w}\hbar t\right)\,,
\ee
where $\tilde w_1$ and $\tilde w_2$ are constants.

The Wigner function provides the energy spectrum by means of the \textit{stargenvalue} equation
\be\label{*gen}
H*W=E\,W\,.
\ee
The Hamiltonian acts diagonally when
\be\label{w0}
w_0=\frac{i\hbar}{2} w_3,
\ee
or alternatively when $w_0=-\frac{i\hbar}{2} w_3$. By definiteness, we choose option \eqref{w0}. Therefore the Wigner function splits into two sectors,
\be
W_+=\frac{i\hbar}{2}w_3+ \pi_\lambda\, w_2 + \lambda\pi_\lambda\, w_3\,,\qquad W_-=\lambda\, w_1\,, 
\ee
such that,
\be
H*W_{\pm}= E_{\pm}W_{\pm}\,,\qquad E_{\pm}=\pm \frac{\hbar\omega}{2}\,. 
\ee

In order to have $\langle 1\rangle=1$, we should normalize $w_3=1$ and the meaning of the constant $w_1$ and $w_2$ is obtained from the expectation values 
\be
\langle\lambda\rangle=w_2\,,\qquad  \langle\pi_\lambda\rangle=-w_1\,.
\ee
Hence we can write,
\be
W_+=\frac{i\hbar}{2}+ \pi_\lambda\, \langle\lambda\rangle + \lambda\pi_\lambda\,\,,\qquad W_-=-\lambda\langle\pi_\lambda\rangle\,, 
\ee
The analogue of the constraints \eqref{qmc} $\pi_\lambda*W=\tilde w_1=0$ implies $\langle\pi_\lambda\rangle=0$ for all $t$.

\section{Conclusions and final remarks}\label{sec:con}

The goal of our article has been to propose a new fermionic counterexample to the Dirac conjecture, which complements the many bosonic cases already existing in the literature (see \eg \cite{Henneaux:1992ig,Cawley1979,Pons:2004pp,Earman:2003cy}). 

Here, a direct test shows that the secondary first-class constraint does not generate a gauge symmetry, and hence the Dirac conjecture fails. We have shown that there is a single gauge symmetry generator, the Castellani chain \eqref{Castellani}, and that the number of gauge fixing conditions and the number of  FCC do not need to match to get rid of all arbitrary functions of time: it suffices to impose a gauge fixing condition conjugate to the primary FCC. Thus the system propagates one degree of freedom. Assuming the Dirac conjecture without verification would lead us to impose an additional constraint, wrongly referred to as ``gauge fixing", and there would be no propagating degrees of freedom.

The gauge fixed model is equivalent to the fermion harmonic oscillator, with the secondary first-class being imposed by means of a boundary condition, which is preserved for all time from the first-order nature of the system. Quantization does not present problems, as suggested in \cite{Henneaux:1992ig} for scenarios where the Dirac conjecture is invalid. Here, the secondary first-class constraint can be imposed as a condition that ``prepares" the system in the positive energy stationary state. Quantization problems of this kind of system result from the lack of definition of the Poisson bracket in odd-dimensional subspaces of the phase space. Indeed, if an odd number of phase space directions are removed algebraically, the Poisson bracket will not be correctly defined in odd-dimensional induced subspaces. However, it is perfectly possible, as it happens here, that setting secondary FCC as boundary conditions, the system will not evolve along those constrained directions, \ie the system will evolve naturally in a sub-manifold of the initial phase space.  

Our result suggests a more general alternative to the treatment of secondary first-class constraint, as boundary conditions preserved in the volution generated by the total Hamiltonian, instead of being imposed as algebraic restrictions on the phase space. This has interesting implications in the quantization scheme, which we shall formalize in a subsequent publication.

\section*{Acknowledgements}
We warmly thank discussions with Per Sundell, Francesco Toppan and, in particular, Jorge Zanelli. This work was partially funded by grant FONDECYT 1220862.

\bibliographystyle{unsrt}

\end{document}